\documentclass[pra,superscriptaddress,twocolumn]{revtex4-1}

\usepackage[pdftex]{graphicx}
\usepackage{graphicx, xcolor}
\usepackage{amsmath,amsthm,amssymb}
\usepackage{amsmath}
\usepackage{color}

\definecolor{darkGreen}{RGB}{0,110,0}
\definecolor{darkBlue}{RGB}{0,0,130}
\usepackage[colorlinks,citecolor=darkGreen,linkcolor=darkBlue,urlcolor=blue,hyperindex]{hyperref}

\newcommand{\mytitle}{Dynamics of quantum coherence in many-body localized systems}

\begin{document}

\title{\mytitle}

\author{Jin-Jun Chen}
\affiliation{Tianjin Key Laboratory of Quantum Optics and Intelligent Photonics, School of Science, Tianjin University of Technology, Tianjin 300384, China}
\email{chenjinjun@email.tjut.edu.cn}

\author{Kai Xu}
\affiliation{Tianjin Key Laboratory of Quantum Optics and Intelligent Photonics, School of Science, Tianjin University of Technology, Tianjin 300384, China}

\author{Li-Hang Ren}
\affiliation{College of Physics and Hebei Key Laboratory of Photophysics Research and Application, Hebei Normal University,
Shijiazhuang, Hebei 050024, China}

\author{Yu-Ran Zhang}
\affiliation{School of Physics and Optoelectronics, South China University of Technology, Guangzhou 510640, China}
\email{yuranzhang@scut.edu.cn}

\author{Heng Fan}
\affiliation{Institute of Physics, Chinese Academy of Sciences, Beijing 100190, China}
\affiliation{Beijing Academy of Quantum Information Sciences, Beijing 100193, China}
\affiliation{Mozi Laboratory,Henan Academy of Sciences, Zhengzhou, Henan 450001, China}
\email{hfan@iphy.ac.cn}

\date{\today}

\begin{abstract}
We demonstrate that the dynamics of quantum coherence serves as an effective probe for identifying dephasing, which is a distinctive signature of many-body localization (MBL). Quantum coherence can be utilized to measure both the local coherence of specific subsystems and the total coherence of the whole system in a consistent manner. Our results reveal that the local coherence of small subsystems decays over time following a power law in the MBL phase, while it reaches a stable value within the same time window in the Anderson localized (AL) phase. In contrast, the total coherence of the whole system exhibits logarithmic growth during the MBL phase and reaches a stable value in the AL phase. Notably, this dynamic characteristic of quantum coherence remains robust even with weak interactions and displays unbounded behavior in infinite systems. Our results provide insights into understanding many-body dephasing phenomena in MBL systems and reveal a novel feasible method for identifying and characterizing MBL phases in experiments.
\end{abstract}


\maketitle

\section{Introduction}

The statistical mechanical description of closed quantum systems suggests that subsystems, containing only a small fraction of the degrees of freedom, typically thermalize by utilizing the remainder of the system as a heat reservoir and lose memory of their initial conditions~\cite{PhysRevA.43.2046,PhysRevE.50.888,rigol2008thermalization,gogolin2016equilibration}. However, Anderson localization (AL)~\cite{anderson1958absence} and many-body localization (MBL)~\cite{basko2006metal} have been widely shown to violate this paradigm. In addition, the phenomena of weak violation, including prethermalization~\cite{PhysRevLett.93.142002,Gring2012,PhysRevX.9.021027} and many-body scars~\cite{PhysRevE.50.888,Bernien2017,Turner2018,Serbyn2021,PhysRevLett.132.020401}, have recently attracted much attention. AL is caused by elastic scattering of random impurities in the absence of interactions, resulting in localization of all single-particle states, which has been directly observed~\cite{wiersma1997localization, billy2008direct, roati2008anderson}. With regard to the interplay between disorder and interacting particles, MBL characterizes the phenomenon that violates thermalization, with its hallmark characteristics being observed in various experimental setups~\cite{kondov2015disorder,schreiber2015observation,bordia2016coupling,alvarez2015localization,smith2016many,xu2018emulating,guo2021observation,PhysRevResearch.3.033043}. Recent theoretical and experimental studies have shown that even without the random component, such as Stark MBL with a linear Zeeman field, a system can enter into an MBL phase~\cite{PhysRevLett.122.040606,PhysRevB.103.L100202,Evert2019From,PhysRevB.101.174204,Bhakuni2020,PhysRevLett.130.120403,PhysRevB.104.205122,PhysRevLett.127.240502,Morong2021}. AL and MBL phases share several key features distinguishing them from ergodic phases, including the absence of the transport of particles and energy~\cite{anderson1958absence,basko2006metal}, the preservation of some memory of local initial conditions, the failure of the eigenstate thermalization hypothesis~\cite{PhysRevB.82.174411}, the existence of eigenstates with area-law entanglement~\cite{RevModPhys.82.277,PhysRevLett.113.107204}, and energy level statistics following Poisson distribution~\cite{PhysRevB.82.174411,PhysRevB.93.041424,PhysRevB.97.035163,PhysRevLett.122.180601}. A comprehensive description can be found in the reviews in ~\cite{nandkishore2015many,RevModPhys.91.021001}.

However, the MBL phase possesses distinctive features that qualitatively distinguish it from the AL phase. This is because while disorder prevents particle scattering, quantum correlations can still be transported when the interaction is turned on. Identifying these distinguishing features of MBL is crucial, and various tools of quantum information theory have proven effective for observing this non-triviality. The dynamics of large-block entanglement after a global quench has been discussed~\cite{vznidarivc2008many,bardarson2012unbounded,PhysRevB.85.094417,PhysRevLett.110.260601,PhysRevB.90.184202,PhysRevB.93.205146,PhysRevB.95.024202}, and the consistent result is that in the AL phase, entanglement entropies for bipartitions tend to reach a plateau, whereas in the MBL phase they exhibit slow logarithmic, but unbounded, growth~\cite{vznidarivc2008many,bardarson2012unbounded,PhysRevLett.110.260601}, which was directly observed experimentally using quantum state tomography~\cite{xu2018emulating}. In small subsystems, the joint state of two subsystems generally becomes a mixed state rather than a pure state, making its entanglement impossible to measure by von Neumann entropy. The dynamics of two-site entanglement, quantified by concurrence, shows a contrasting behavior compared to bipartite entanglement across the whole system: in the MBL phase it decays over time, following a power-law, while in the AL phase it tends towards a plateau~\cite{PhysRevB.94.214206}. Then, quantum mutual information based on von Neumann entropy also exhibits unbounded spreading within the MBL phase~\cite{PhysRevLett.118.016804}. In addition, spin-noise spectroscopy has also been proved to be effective for distinguishing MBL from AL~\cite{PhysRevB.92.180205}. Essentially, the interaction induces dephasing~\cite{PhysRevLett.110.260601} within the MBL phase by reducing the off-diagonal elements in the reduced density matrices for small subsystems, thus conferring non-trivial dynamical behaviors on measures such as entanglement entropy, concurrence, and quantum mutual information. Therefore, it is important to find an appropriate probe that better reflects the many-body dephasing properties under existing experimental conditions, and we demonstrate that quantum coherence serves as such a probe.

In closed quantum systems, quantum coherence arises from the superposition principle and is responsible for quantum fluctuations and correlations. It is primarily characterized by the off-diagonal terms of the density operator, which makes it a natural probe to explore many-body dephasing phenomena. Recently, a rigorous framework for quantum coherence as a resource was established~\cite{baumgratz2014quantifying}, with measures such as relative entropy of the coherence and the $l_{1}$  norm of the coherence being recognized as legitimate quantifiers. Further, quantum coherence has been incorporated into the general framework of quantum resource theories~\cite{RevModPhys.89.041003,RevModPhys.91.025001,PhysRevA.109.022403}. Quantum coherence has been proven to be an effective tool for addressing various quantum many-body problems, including investigating different types of quantum phase transitions~\cite{chen2016coherence,PhysRevA.98.012303} and studying the localization transition between the ergodic phases and MBL phases~\cite{styliaris2019quantum,PhysRevA.99.052354,dhara2020quantum}. We employ the dynamics of quantum coherence to examine the distinction between the MBL phase and the AL phase. In multi-particle systems, the total coherence has contributions from both local coherence arising within each defined subsystem and intrinsic coherence emerging from inter-subsystem correlations~\cite{PhysRevLett.116.150504}. Another advantageous feature of quantum coherence is that it does not require system partitioning, and measures, such as the relative entropy of the coherence and the $l_{1}$ norm of the coherence can directly quantify the total coherence of the whole system while also serving as valid indicators when we focus on the local coherence of subsystems specifically. Unlike entanglement, for which distinct measures are needed for bipartite entanglement versus two-site entanglement, quantum coherence allows us to consistently measure both local coherence within individual subsystems and the total coherence of the whole system in a unified manner, thereby facilitating exploration of the relationship between the local and global aspects of coherent behavior. Furthermore, it was shown experimentally that quantum coherence can be measured directly with interference fringes~\cite{PhysRevLett.118.020403}, which is easier under existing experimental conditions compared to quantum state tomography.

In this paper, we investigate the dynamics of quantum coherence in the interacting fermion model. We begin by initializing the system in a N\'eel state and adopt the $l_{1}$ norm of coherence as a measure of quantum coherence. We explore the dynamical behavior of quantum coherence in both the nearest two-site subsystem and the whole system, demonstrating its ability to distinguish among the MBL phase, non-interacting AL phase, and non-local ergodic phase. Subsequently, we present the distribution of quantum coherence within both the whole system and various dimensional subsystems. Furthermore, by considering weak interactions and varying chain lengths, we establish that quantum coherence undergoes unbounded slow changes within the MBL phase. Finally, we supplement our findings from alternative perspectives by presenting results for two other initial states and the relative entropy of coherence.

\section{model}

We employ an interacting-fermion chain with disordered potential, described by the Hamiltonian:
\begin{equation}
 H=J\sum_{l=1}^{N}\left(c^\dagger_l c_{l+1} + c^\dagger_{l+1} c_{l}\right) + \Delta\sum_{l=1}^N n_l n_{l+1}+\sum_{l=1}^Nh_l n_l,
 \label{eq:Hamiltonian}
\end{equation}
where $c^{(\dagger)}_l$ annihilates (creates) a fermion on site $l$, and $n_l=c^\dagger_l c_l$ represents the local number operator. The single-particle basis states can be constructed from the vacuum $|0\rangle$, and the excited state is $|1\rangle = c^\dagger |0\rangle$. Here, $J$ and $\Delta$ denote the hopping strength and interacting strength respectively, and the local disordered potentials $h_l$  are taken from a uniform random distribution in $\left[-\delta h,\delta h\right]$. As the disorder strength increases, there are two distinct types of transitions from the ergodic phase to the localized phase. First, in the noninteracting case with $\Delta=0$, the AL transition occurs at $\delta h_c=2$~\cite{Aubry1980}. Second, in the interacting case with $\Delta\neq 0$, the  MBL transition is observed where its phase transition point depends on $J$ and $\Delta$ values chosen. For instance, the choice $J=1/2$, $\Delta=1$  is believed to have a transition at $\delta h_c=3.5\pm 1$~\cite{PhysRevB.82.174411,PhysRevLett.115.046603,PhysRevB.91.081103}. For our model, we infer that the phase transition occurs at $\delta h_c=5.8\pm 0.06$ by utilizing level statistics (see the Appendix).

In order to detect signatures of quantum coherence in the AL phase and MBL phase, we perform a global quench from an initial state far from equilibrium, then follow the time evolution governed by the Hamiltonian ~\eqref{eq:Hamiltonian}. In general, we choose the initial state to be the N\'eel state $|\psi(t=0)\rangle=|1010\cdots1010\rangle$, and the energy of this configuration is equivalent to an infinite temperature thermal state since its expected value of the Hamiltonian is zero~\cite{xu2018emulating}. To further demonstrate the effectiveness of quantum coherence, two additional initial states are considered: (1) another incoherent initial state $|\psi(t=0)\rangle=|11\cdots1100\cdots00\rangle$ with all excitations confined to the left half-chain and corresponding energy equivalent to an infinite temperature thermal state~\cite{xu2018emulating}, and (2) a maximally coherent state (introduced in the next section).

We study the above-mentioned model ~\eqref{eq:Hamiltonian} using the QUTIP packages~\cite{QuTiP2012,QuTiP2013}. In the numerical calculations, the hopping strength $J=1$ is fixed, and $1000$ disorder realizations are used to perform the disorder averaging (except for the saturation values in the inset of Fig.~\ref{Coh-N:fig} below). Unless otherwise indicated, we present the basic results with the settings that the system has $N=12$ sites initialized at the N\'eel state and interaction strength $\Delta=1$ for MBL phase.

\section{Measures of quantum coherence}

\begin{figure}
  \includegraphics[width=0.9\columnwidth]{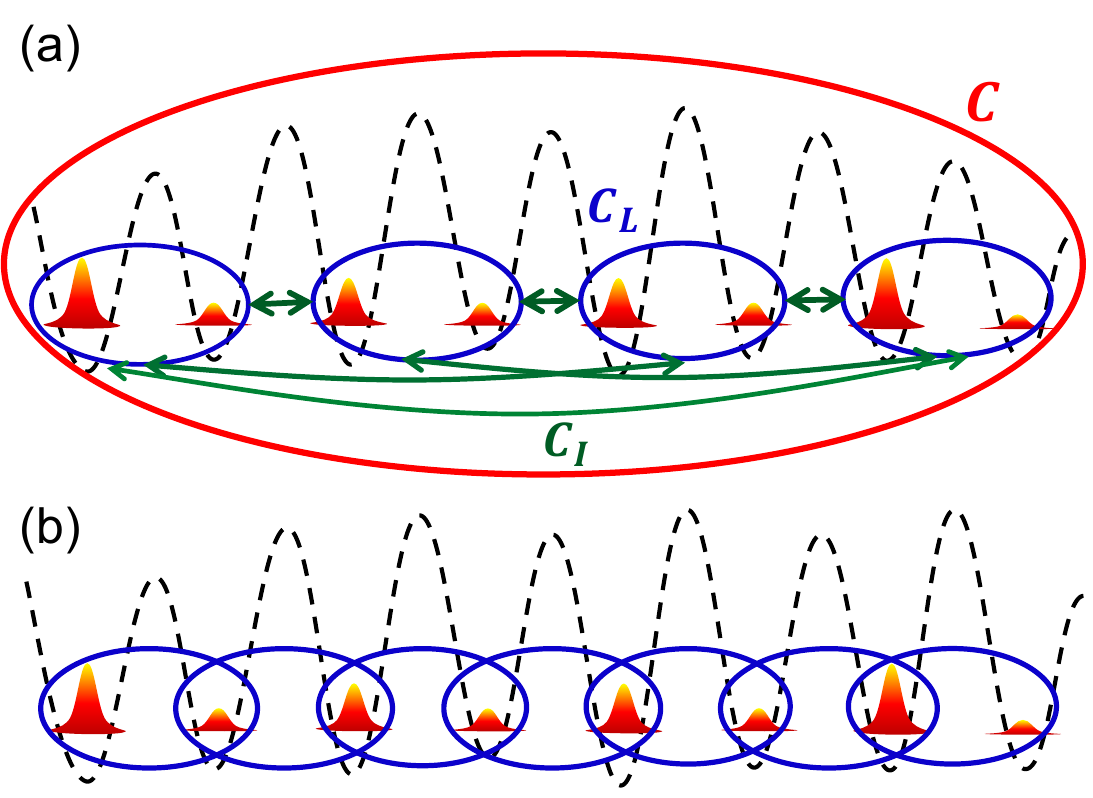}
  \caption{Quantum coherence in an interacting-fermion chain with disordered potential. (a) The total coherence $C$ is contributed by local coherence $C_L$ on subsystems and intrinsic coherence $C_I$. Here, the system is divided into several $n=2$ subsystems uniformly; in fact, the system can be divided arbitrarily. (b) Taking the subsystem dimension $n=2$ as an example, the selection method of all the nearest-neighbor $n$-site subsystems in the system.}
  \label{Illustration:fig}
\end{figure}

Based on the measurement scheme proposed in Ref.~\cite{baumgratz2014quantifying}, the concept of quantum coherence is based on a given basis under which all diagonal density matrices form a set of incoherent states. Incoherent operations are required to map this set onto itself. Generally, for a $d$-dimensional Hilbert space with a particular basis, $\{|i\rangle\}_{i=1,\cdots,d}$, a $d$-dimensional maximally coherent state can be specified as
\begin{equation}
|\Psi_{d}\rangle = \frac{1}{\sqrt{d}} \sum_{\substack{i}}^d |i\rangle,
 \label{MaxCoh}
\end{equation}
which allows the definition of a coherence unit to normalize all measures. Note that the maximally coherent states are not unique, and the complete set of these states is given in Ref.~\cite{PhysRevA.93.032326}.

The central requirement of the coherence measure is that the coherence should not increase under incoherent operations. The $l_1$ norm of the coherence expressed as the sum of off-diagonal elements has been proven to be a valid measure of coherence for a given basis,

\begin{equation}
 C_{l_1}(\rho)=\sum_{\substack{i,j \\ i \neq j}}|\rho_{i,j}|.
 \label{Cl1}
\end{equation}
Another universally valid measure is the relative entropy of the coherence,
\begin{equation}
C_{\text{rel.ent}}(\rho) = S(\rho_{\text{diag}}) - S(\rho),
 \label{Cren}
\end{equation}
where $S$ represents the von Neumann entropy and $\rho_{\text{diag}}$ is obtained from $\rho$ by removing all its off-diagonal entries.

The notions of local coherence and intrinsic coherence were introduced in Ref.~\cite{PhysRevLett.116.150504}. We can now study the properties of the total coherence of the system and the local coherence of a specific subsystem in many-body dynamics. Figure~\ref{Illustration:fig}-(a) illustrates that the total coherence of the whole system has contributions from the local coherence in subsystems and the intrinsic coherence between subsystems using a specific example. Since the measure of coherence depends on the basis, we choose the tensor product basis, formed by the vacuum state $|0\rangle$ and the excited state $|1\rangle$ of the fermion, as the preferred basis throughout the paper. Then, we calculate the total coherence of the whole system and the local coherence of any dimensional subsystems, uniformly using the coherence measures in Eqs.~\eqref{Cl1} and ~\eqref{Cren}, where the local coherence of the subsystem uses the corresponding reduced density matrix. For the local coherence of the $n$-site subsystem ($2\leq n < N$), we averaged all the nearest-neighbor $n$-site subsystems, [see Fig.~\ref{Illustration:fig}-(b)]. The $n$-site local coherence can be expressed as
\begin{equation}
C(\rho_{n}) = \frac{1}{N+1-n}\sum_{\substack{i=1}}^{N+1-n}C(\rho_{[i,i+1,\cdots,i+n-1]}),
\label{Coherence_n}
\end{equation}
where $\rho_{[i,i+1,\cdots,i+n-1]}$ is the reduced density matrix describing the subsystem formed by the nearest-neighbor $[i,i+1,\cdots,i+n-1]$ sites, and $C(\bullet)$ is the unified representation of $C_{l_1}(\bullet)$ and $C_{\text{rel.ent}}(\bullet)$. Then the total coherence of the whole system has the unified expression $C(\rho_{N}) = C(\rho_{n=N}).$

In many cases, the single-site density matrix is non-trivial for coherence, and the single-site local coherence can be expressed as
\begin{equation}
C(\rho_{n=1}) = \frac{1}{N}\sum_{\substack{i=1}}^{N} C(\rho_{[i]}).
\label{}
\end{equation}

For the relative entropy of the coherence, the disorder-averaged coherence is $\langle C_{rel.ent}(\rho_n) \rangle$ with $\langle\bullet\rangle$ indicating the disorder average. However, we usually normalize the $l_1$ norm of coherence with the maximally coherent state ~\eqref{MaxCoh}, then the disorder-averaged normalized coherence is given by,
\begin{equation}
\langle C_{l_1}^N(\rho_n) \rangle= \frac{1}{2^n-1}\langle C_{l_1}(\rho_n) \rangle, (1\leq n \leq N).
\label{l1_normalized}
\end{equation}

\section{Quantum coherence in disordered fermion chain}

In this section, we consider the initial state to be the N\'eel state, and the one-site reduced density operators are diagonal during the evolution. Quantum coherence remains at zero. Thus, we focus on the non-trivial cases with $n\geq2$. The coherence measure adopts the $l_1$ norm of the coherence by disorder-averaged normalization in Eq.~\eqref{l1_normalized}.

\begin{figure}
  \includegraphics[width=0.98\columnwidth]{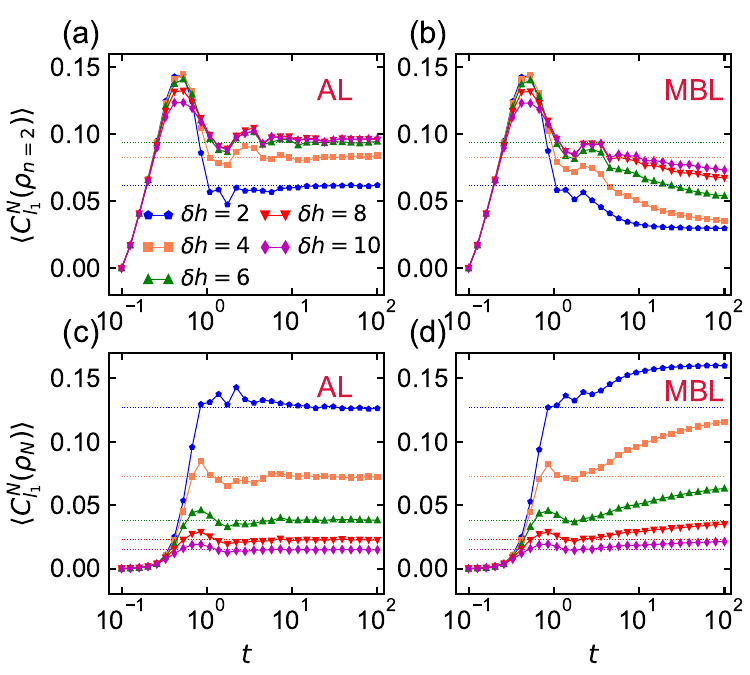}
  \caption{Time evolution of coherence in the AL model ($\Delta=0$) and MBL model ($\Delta=1$). The system has $N=12$ sites initialized at the N\'eel state, and values of $\delta h$ are chosen accordingly. (a) and (b) show the local coherence for $n=2$ subsystems, and (c) and (d) show the total quantum coherence. The color coding and markers are the same in all four panels.}
  \label{Coh-AL-MBL:fig}
\end{figure}

\subsection{Coherence as a probe to detect the  uniqueness of MBL}

\begin{figure}
  \includegraphics[width=0.98\columnwidth]{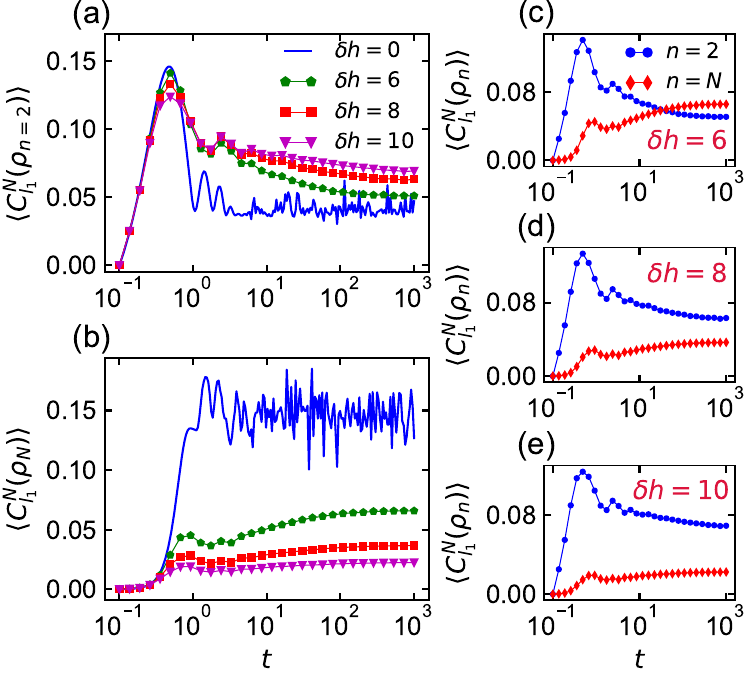}
  \caption{Time evolution of coherence in the absence of disorder (solid line) and strong disorder (line with markers). The system has $N=12$ sites initialized at the N\'eel state, and interaction strength $\Delta=1$. (a) and (b) show the time evolutions of the two-site local coherence and the total quantum coherence at the different disorder strengths $\delta h$ indicated, and have the same color coding and markers. (c)-(e) show a comparison between two-site local coherence and the total coherence with disorder strengths $\delta h=6$, $\delta h=8$, and $\delta h=10$, respectively; the color coding and markers are the same.}
  \label{Coh-MBL:fig}
\end{figure}

\begin{figure*}
  \includegraphics[width=1.98\columnwidth]{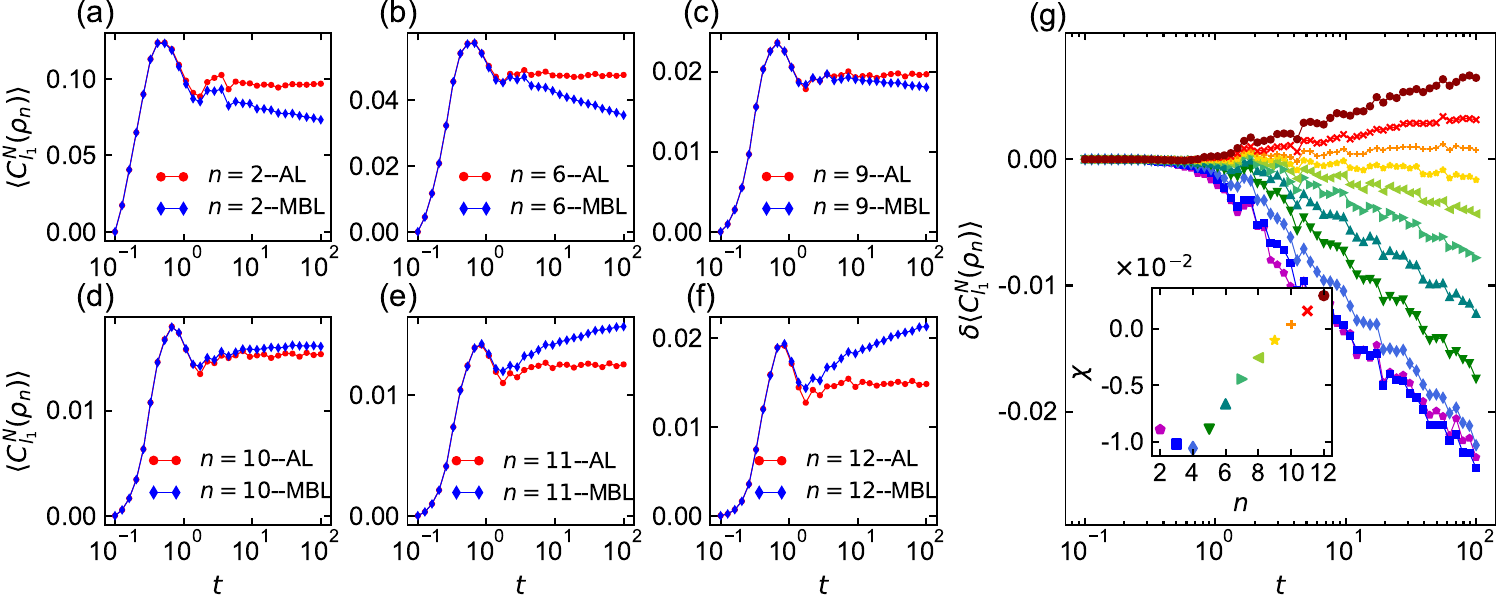}
  \caption{Dynamics of quantum coherence in the AL phase ($\Delta=0$) and MBL phase ($\Delta=1$) for subsystems of different dimensions $n$. The system has $N=12$ sites initialized at the N\'eel state, with strong disorder $\delta h=10$. (a)-(c) correspond to subsystem dimensions of $n=2$, $n=6$, and $n=9$, respectively, where coherence remains in the AL phase and the local coherence power law decays in the MBL phase. (d)-(f) correspond to subsystem dimensions of $n=10$, $n=11$, and the whole system, respectively, where coherence remains in the AL phase and logarithmically grows in the MBL phase. (g) shows the difference between $\Delta=1$ and $\Delta=0$ for each $n$; and the inset shows the slope of the data in the main plot for $t>10$ for each $n$, where the color coding and markers are uniform across the main plot and the inset.}
  \label{Coh-n:fig}
\end{figure*}

We consider a system with $N=12$ sites and study the evolution of local coherence for $n=2$ subsystems and the total coherence for the whole system under different disorder strengths in the AL model ($\Delta=0$) and the MBL model ($\Delta=1$). Figures~\ref{Coh-AL-MBL:fig}-(a) and ~\ref{Coh-AL-MBL:fig}(b) show the dynamics of local coherence for $n=2$ subsystems in the AL model and MBL model, respectively. The results show that for a short time $t\leq1$, the local coherence behaves exactly the same in both cases, experiencing a rapid increase to a maximum value and then decreasing. As the disorder strength increases, the maximum of the local coherence decreases, indicating that this response is independent of the interaction. Then, in the AL phase, the local coherence reaches a stable value and remains frozen. The stability values increase as the disorder strength increases, but for larger disorder strengths $\delta h=6,8,10$, the stability values are very close to each other and difficult to distinguish. At the same time in the MBL phase, local coherence exhibits a power-law decay. As the disorder strength increases, the decay process becomes slower and more persistent.

Figures~\ref{Coh-AL-MBL:fig}-(c) and ~\ref{Coh-AL-MBL:fig}(d) show the results of the total coherence of the whole system in the AL model and MBL model, respectively. Similarly, within a short time period of $t\leq1$, the total coherence behaves identically in both cases, independent of interactions. In the AL phase, the total coherence reaches a stable value. The differences lie in two aspects: (1) as the disorder strength increases, the saturation value of the total coherence decreases in the AL phase; (2) in the MBL phase, the total coherence exhibits logarithmic growth, and as the disorder strength increases, the growth becomes slower and lasts longer. In the MBL phase, the power-law decay of the local coherence and the logarithmic growth of the total coherence are consistent with the entanglement results, which exhibit a decay of the concurrence~\cite{PhysRevB.94.214206} and a logarithmic growth of the half-chain entanglement~\cite{vznidarivc2008many,bardarson2012unbounded}. This is because during the dynamical evolution, interaction-induced energy differences enter off-diagonal terms of the reduced density matrix for small subsystems, leading to their reduction~\cite{PhysRevLett.110.260601}. The dephasing gives rise to non-trivial entanglement dynamics, where quantum coherence defined by summing off-diagonal elements can most directly reflect the characteristics of this many-body dephasing process and serve as an effective indicator for detecting MBL features.

We provide a qualitative explanation for the behavior of the local coherence in small subsystems and the total coherence in the whole system. The total coherence encompasses not only local coherence within small subsystems but also the intrinsic coherence arising from coherence between distinct subsystems. In the AL phase, as the disorder strength increases, more coherence is stored in the small subsystems, leading to an increase in the local coherence within them. Then the intrinsic coherence decreases significantly, leading to a decrease in the total coherence. In the MBL phase, dephasing caused by interactions allows correlations to propagate continuously, resulting in slow decay of the local coherence stored in small subsystems. Then the intrinsic coherence between the subsystems increases, which leads to a slow logarithmic increase in the total coherence of the whole system.

Next, we explore the evolution of coherence in the presence of strong disorder and in the absence of disorder, with the interaction strength $\Delta=1$. Figures~\ref{Coh-MBL:fig}-(a) and~\ref{Coh-MBL:fig}(b) show the two-site local coherence and the total coherence of the whole system respectively, with time extended up to $t=1000$. When the disorder strength is $\delta h=0$, the system enters the ergodic phase; in the presence of strong disorder $\delta h=6,8,10$, the system enters the MBL phase. Combining the two cases, we find that in the ergodic phase, the time evolution of the coherence can be roughly divided into two stages: first, a period of rapid increase is driven by the hopping term, leading to a maximum in a short time frame. Second, the local coherence undergoes a sharp drop to reach its minimum and oscillates nearby, while the total coherence continues to oscillate around its maximum. For finite-size systems in the MBL phase, the time evolution of coherence can be approximately divided into four stages: The first stage is also a period of rapid increase, caused by hopping terms. In the second stage, there is a subsequent rapid decrease (both these stages are consistent with AL phase and independent of interactions). In the third stage, the local coherence exhibits a slow power-law decay, while the total coherence exhibits a slow logarithmic growth. Finally, if sufficient time is given in the fourth stage, the saturation of coherence is reached. As the disorder strength increases, the duration of the third phase is extended, delaying the entry into the fourth phase. Moreover, the saturation value increases for the local coherence but decreases for the total coherence with increasing disorder. Thus, in the ergodic phase, the local coherence rapidly diminishes and spreads throughout the system, whereas either AL or MBL preserves the local coherence.

In Figs.~\ref{Coh-MBL:fig}-(c)-\ref{Coh-MBL:fig}(e), we visually compare the results of two-site local coherence and total coherence with time for disorder intensity $\delta h=6,8,10$, respectively. The results show that the interaction effect on the time range is the same, when the local coherence power law decays, the total coherence also increases logarithmically, and the time of entering the saturation period is also consistent [Fig.~\ref{Coh-MBL:fig} (c)]. Note that we present quantum coherence $\langle C_{l_1}^N(\rho_n) \rangle$ normalized by the maximally coherent state here, which is convenient for showing the trend of local coherence and total coherence in the same figure. In fact, the coherence of the total system $\langle C_{l_1}(\rho_{n=12}) \rangle$ is much larger than local coherence $\langle C_{l_1}(\rho_{n=2}) \rangle$.

\subsection{Coherence distribution for different subsystems in the evolution}

In the previous section, we examined the cases of local coherence of the smallest non-trivial subsystems and total coherence of the whole system. Quantum coherence allows us to investigate local coherence for subsystems of arbitrary dimension. Therefore, this section focuses on the time evolution of subsystems with dimensions ranging from $n=2$ to $11$, up to $N=12$ for the full system, as well as exploring the distribution of coherence in subsystems with different dimensions. The disorder strength is $\delta h=10$. Fig.~\ref{Coh-n:fig} (a--f) show a comparison of the time evolution of quantum coherence in AL and MBL phases for different $n$. It is observed that quantum coherence reaches to a constant value in all cases within the AL phase. In contrast, in the MBL phase, the quantum coherence decays as a power-law for $n$ exceeding $N/2$ up to $n=9$. However, when $n=10,11,12$, quantum coherence transforms into logarithmic growth. Based on these observations at different values of $n$, it appears that as the dimensionality of the subsystem increases within the MBL phase, the quantum coherence undergoes a process involving a decreasing power-law decay amplitude followed by a transition to a logarithmic growth and finally an increasing logarithmic growth amplitude. Consequently, we introduce a difference for each $n$, denoted as,
\begin{equation}
\delta\langle C_{l_1}^N(\rho_n) \rangle= \langle C_{l_1}^N(\rho_n) \rangle_{\Delta=1} - \langle C_{l_1}^N(\rho_n) \rangle_{\Delta=0},
\end{equation}
its variations for all values of $n$ are depicted in Fig.~\ref{Coh-n:fig}-(g). Notably during the AL saturation period ($t>10$), it can be observed that there a linear relationship exists between the difference $\delta\langle C_{l_1}^N(\rho_n) \rangle$ and $\log_{10} t$.
We define a slope representing this linear relation, labeled by
\begin{equation}
\chi(n)= \frac{d \delta\langle C_{l_1}^N(\rho_n) \rangle}{d(\log_{10} t)} \bigg|_{t>10}.
\end{equation}
The corresponding values for each subsystem dimension $n$ are shown in the inset. Remarkably,  for cases where $n>4$, it becomes evident that slope $\chi$ steadily increases with respect to increasing values of $n$. This observation confirms that in the MBL phase and for $n>4$, as $n$ increases, quantum coherence surely undergoes a progression from power-law decay amplitude reduction to increasing logarithmic growth amplitude. This transformation is a continuous behavior rather than a transition. When the subsystem dimension increases almost to the entire system dimension, the local coherence of the subsystem is close to the total coherence of the whole system and exhibits dynamic characteristics that are consistent with the total coherence of the whole system.

\subsection{Unbounded change of coherence}

\begin{figure}
  \includegraphics[width=0.9\columnwidth]{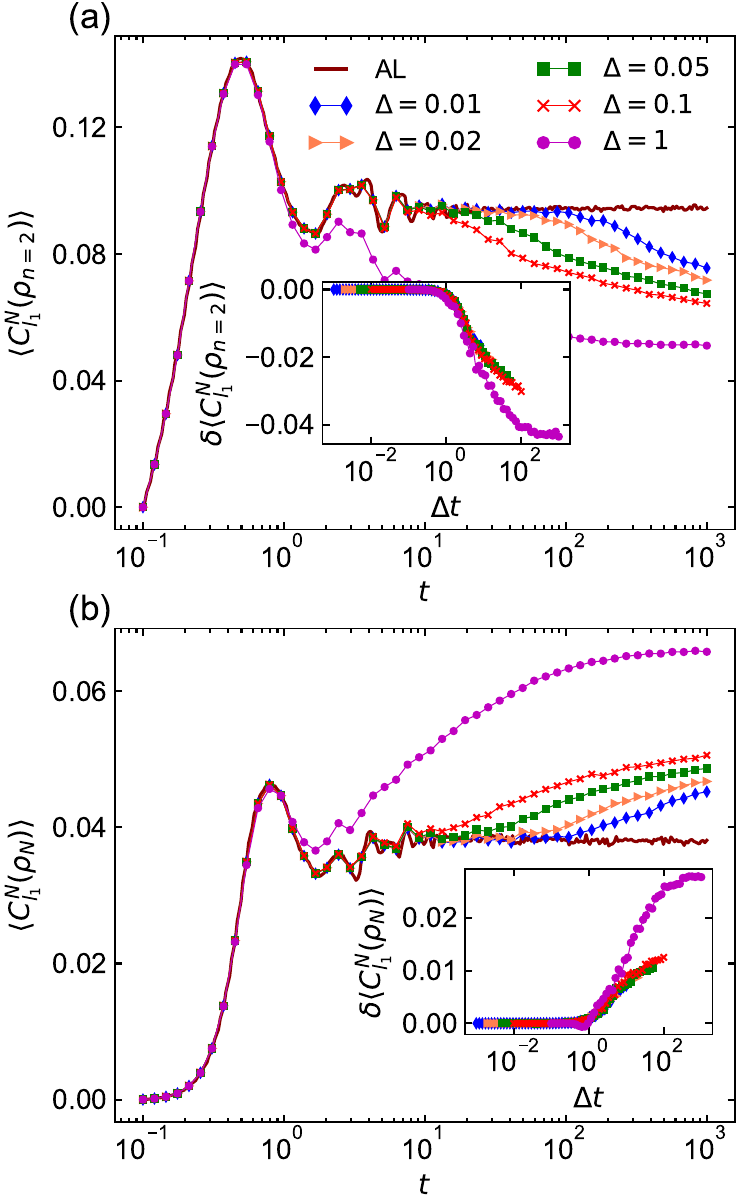}
  \caption{Dynamics of coherence for different interaction strengths $\Delta$ indicated. The system has $N=12$ sites initialized at the N\'eel state, and disorder strength $\delta h=10$. (a) shows the subsystem dimension of $n=2$, and (b) shows the whole system, showing that even for very small interaction terms $\Delta$, the dynamics of coherence is substantially different from that in the non-interaction model. The inset shows the same data-subtracted $\Delta=0$ values, but with a rescaled time axis. The color coding and markers are the same in both panels.}
  \label{Coh-Delta:fig}
\end{figure}

The presence of interactions determines whether the system exhibits AL or MBL phenomena. In this section, we focus on comparing the AL phase and MBL phase in the weak interaction limit ($\Delta\ll1$), in which the entanglement has been discussed~\cite{bardarson2012unbounded,PhysRevB.94.214206}. The evolution of local coherence for the $n = 2$ subsystem and total coherence of the whole system under different interaction strengths are represented by Fig.~\ref{Coh-Delta:fig} (a) and (b), with the disorder strength $\delta h=10$. We observe a distinct difference between the curves in the weak interaction limit ($\Delta$ ranges from $10^{-2}$ to $10^{-1}$) and at $\Delta=1$.
For $\Delta=1$, the effects of interaction start to appear after $t>1$, causing a separation from the curve at $\Delta=0$. However, in case of weak interactions, there are two distinct stages after $t>1$: initially agreeing with the $\Delta=0$ curve and indistinguishable from the AL phenomenon; then exhibiting MBL signature characterized by power-law decay in local coherence and logarithmic growth in total coherence. This suggests that in the weak interaction limit, smaller interactions lead to a later onset of the MBL effect.

By plotting the difference between $\langle C_{l_1}^N(\rho_n)_{\Delta}\rangle$ and $\langle C_{l_1}^N(\rho_n)_{\Delta=0}\rangle$ against rescaled time axis $\Delta t$ as shown in inset, we observe that all curves for weak interaction limits collapse onto a single curve where the signature of MBL starts at $\Delta t=1$. These results demonstrate that even very small interactions can lead to MBL within the system. Moreover, for the weak interaction limit, the time at which the MBL effect starts is inversely proportional to the strength of the interaction $t=1/\Delta$.

\begin{figure}
  \includegraphics[width=0.9\columnwidth]{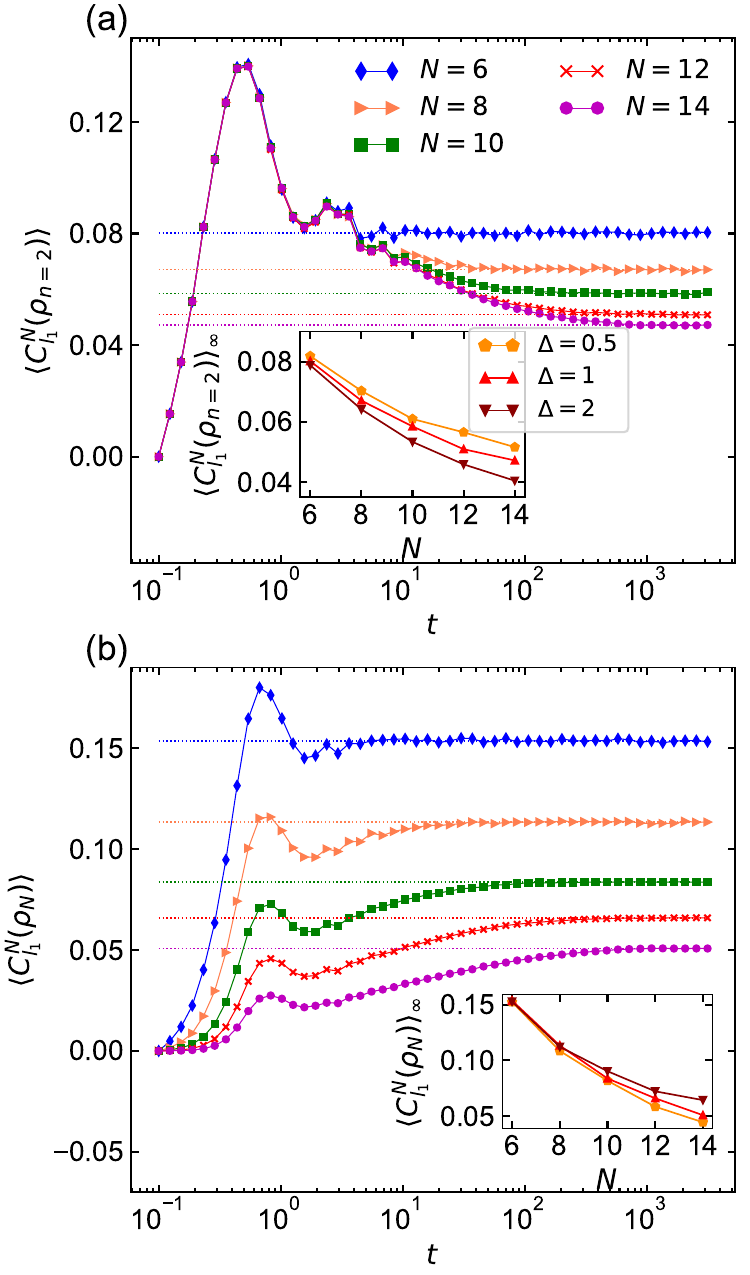}
  \caption{Dynamics of coherence for the different system dimensions $N$ indicated. The initial state is the N\'eel state, and disorder strength is $\delta h=6$. (a) shows a subsystem dimension of $n=2$, and (b) shows the whole system, interaction strength $\Delta=1$. The inset shows the saturation values (averaged over $100$ disorder realisations for $\Delta=0.5$ and $\Delta=2$) of the coherence as a function of $N$ for different interaction strengths. The color coding and markers are the same in both panels.}
  \label{Coh-N:fig}
\end{figure}

From the above discussion, it is evident that in finite systems, quantum coherence reaches a saturation value in the MBL phase after undergoing the period of logarithmic growth (or power-law decay). Subsequently, we investigate the correlation between the saturation values of quantum coherence and system size. Considering that a relatively small disorder strength and a relatively large interaction strength facilitate more rapid saturation of coherence, we set the parameters as $\delta h=6$ and $\Delta=1$. Fig.~\ref{Coh-N:fig} (a) and (b) show the results of local coherence for $n=2$ and total coherence for the whole system, respectively. These results indicate that the time $t_{sat}$ at which the coherence saturates is consistent with the scaling $\log_{10} t_{sat}\sim N$. In addition, for $2$-site local coherence, its saturation value decreases with increasing $N$. In fact, the total coherence will increase in content with larger system sizes. However, we use the normalized $l_1$ norm of coherence by the maximally coherent state, therefore total coherence $\langle C_{l_1}^N(\rho_N) \rangle$ represents the ``coherence level''. The results reveal that as $N$ increases, there is a decrease in the saturation value of ``coherence level''. In insets, the relationship between the saturation value and the size of the system is shown, and in both cases the saturation value is inversely proportional to the system size $N$. Additionally, we present results for $\Delta=0.5$ and $\Delta=2$ in insets of Fig.~\ref{Coh-N:fig} (a) and (b), which also exhibit that the saturation value is inversely proportional to $N$. Based on our observation that larger $N$ requires more time to reach saturation state, it can be anticipated that such change (logarithmic growth or power-law decay) remains unbounded in infinite systems.

\section{Results for other initial states}

\begin{figure}
  \includegraphics[width=0.98\columnwidth]{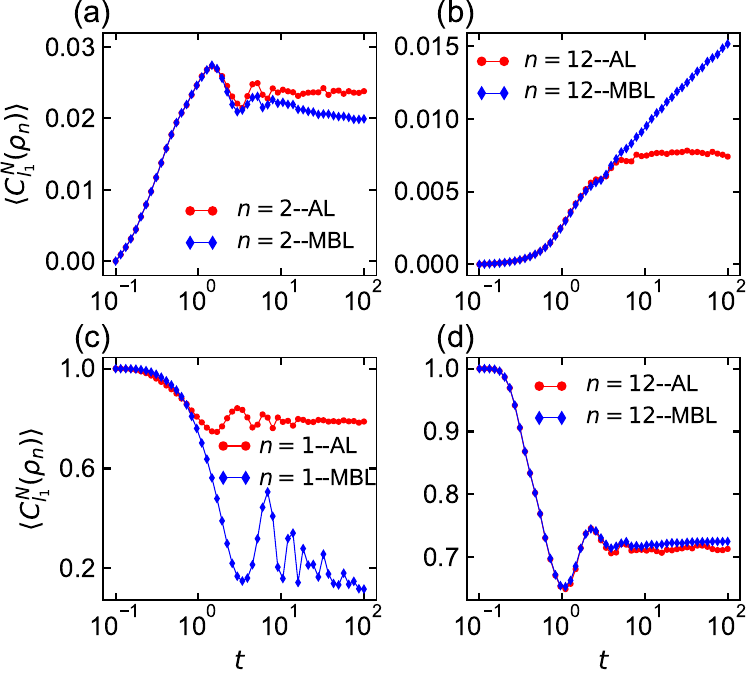}
  \caption{Dynamics of coherence for two different initial states. The system has $N=12$ sites, interaction strength $\Delta=1$, and disorder strength $\delta h=10$. The incoherent initial state $|\psi(t=0)\rangle=|111111000000\rangle$ is shown in (a) and (b) for $n=2$ subsystems and the whole system, respectively. The maximal coherent initial state is shown in (c) and (d) for subsystem dimension $n=1$ and the whole system, respectively.}
  \label{OtherStates:fig}
\end{figure}

\begin{figure*}
  \includegraphics[width=1.98\columnwidth]{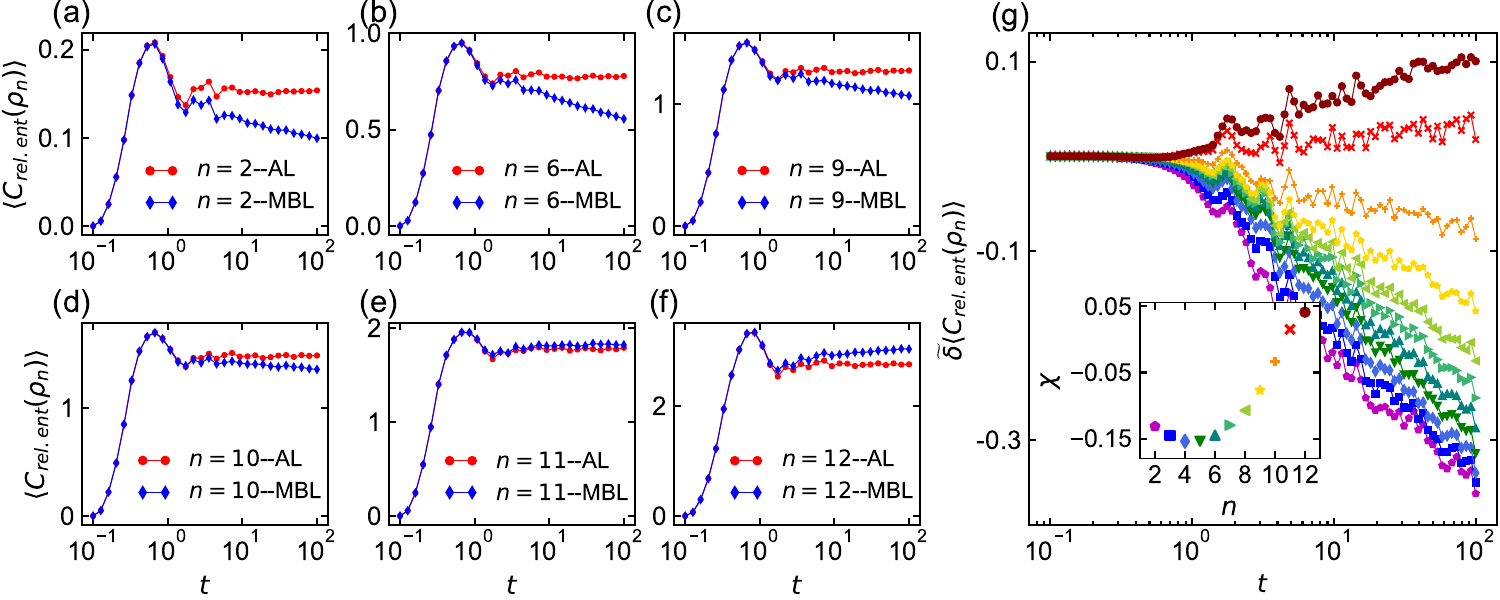}
  \caption{Relative entropy of coherence for subsystems of different dimensions $n$. The system has $N=12$ sites initialized at the N\'eel state, and disorder strength $\delta h=10$. (a)-(f) correspond to $n=2,6,9,10,11,12$, respectively. (g) shows the ``difference'' between $\Delta=1$ and $\Delta=0$ for each $n$, and the inset shows the slopes of the data in the main plot for $t>10$ for each $n$; where the color coding and markers are the same in the main figure and the inset.}
  \label{Rel.Ent.coh:fig}
\end{figure*}

In the above sections we have shown the dynamical features of quantum coherence in the initial N\'eel state. It is well known that the coherent evolution outcome is related to the choice of the initial state. To further substantiate the efficacy of quantum coherence in detecting MBL properties, we chose two distinct types of initial states and examine their evolution under strong disorder $\delta h=10$.

First, we present the coherence results for an incoherent initial state $|\psi(t=0)\rangle=|111111000000\rangle$ with all excitations confined to the left half-chain, then time evolution of $2$-site local coherence and the total coherence are shown in Fig.~\ref{OtherStates:fig} (a) and (b). The results show that the local coherence decays as a power-law in the MBL phase, when entering saturation in the AL phase. However, the total coherence increases logarithmically in the MBL phase, as saturation enters in the AL phase. Compared to the results for the N\'eel state, the time for the quantum coherence to plateau in the AL phase becomes longer due to the Lieb-Robinson bound~\cite{Lieb1972Com}.

In addition, we present the results of employing the maximally coherent state Eq.~\eqref{MaxCoh} as the initial state. In this case, the quantum coherence value of the initial state is $1$, and the single-site reduced density matrix exhibits non-trivial quantum coherence. We show the results of single-site local coherence and the total coherence in Fig.~\ref{OtherStates:fig} (c) and (d). Considering the single-site coherence, after an initial decay, the quantum coherence stabilizes at approximately $0.8$ in the AL phase, but exhibits a significant power-law decay in the MBL phase. For the total coherence, after an initial decay, the quantum coherence stabilizes around $0.7$ in the AL phase while it increases logarithmically in the MBL phase.

We demonstrate the dynamical nature of quantum coherence from various perspectives, encompassing both incoherent initial states and maximally coherent state. For different initial states, the quantum coherence can clearly show the essential difference between the quantum coherence in the AL and MBL phases. These cases collectively reveal a general property:
quantum coherence enters saturation in the AL phase, but exhibits distinct behaviors in the MBL phase. Specifically, the local coherence of the small system decays by a power law, but the total coherence of the whole system increases by a logarithm.

\section{Results for the Relative entropy of the coherence}

In this section, we present results for the relative entropy of coherence using the same parameter settings and initial state as shown in Fig.~\ref{Coh-n:fig}, aiming to compare two different methods for measuring coherence.
We use the unnormalized disorder-averaged relative entropy of coherence $\langle C_{rel.ent}(\rho_n) \rangle$.
Figure~\ref{Rel.Ent.coh:fig} (a--f) shows a comparison of the time evolution of quantum coherence in AL and MBL for various subsystem dimensions $n$.
The difference with the results obtained using the $l_1$ norm of coherence is that the transition from a power-law decay to a logarithmic growth occurs when the subsystem dimension reaches $n=11$ instead of $n=10$. Similarly, we define a ``difference'' as,
\begin{equation}
\tilde{\delta}\langle C_{rel.ent}(\rho_n) \rangle= \frac{\langle C_{rel.ent}(\rho_n) \rangle_{\Delta=1} - \langle C_{rel.ent}(\rho_n) \rangle_{\Delta=0}}{\langle C_{rel.ent}(\rho_n) \rangle_{\Delta=0}|_{t\to\infty}}.
\end{equation}
where values at $t\to\infty$ are obtained by averaging results for $t>10$ in the AL phase; all results of ``difference'' corresponding to $n$ are presented in Fig.~\ref{Rel.Ent.coh:fig} (g), where it follows a simple linear relationship with $\log_{10} t$ during the AL saturation period ($t>10$).
The slope of this linear relation is defined as,
\begin{equation}
\chi(n)= \frac{d \tilde{\delta}\langle C_{rel.ent}(\rho_n) \rangle}{d(\log_{10} t)} \bigg|_{t>10} .
\end{equation}
The insert demonstrates that when $n>5$, there is a steady increase in slope with respect to $n$.
Comparing the results for the relative entropy of coherence with those for the $l_1$ norm of coherence, we find that the two different measures lead to slight differences in details, however, they do not alter the qualitative findings, namely, when the quantum coherence reaches saturation during the time evolution within an AL phase, the local coherence decreases following a power-law while the total system coherence increases logarithmically within an MBL phase.

\section{Conclusions}
\label{Conclusions}

In this paper, we demonstrate that the dynamics of quantum coherence serves as an effective probe to detect many-body dephasing. Both the local coherence of small subsystems and the total coherence of the whole system can effectively distinguish the MBL phase from the AL phase, but exhibit contrasting behaviors. Specifically, the quantum coherence reaches to a constant value within the AL phase in both cases, but decays as a power law for local coherence in small systems and grows logarithmically for total coherence within the MBL phase. This logarithmic growth (or the power-law decay) characteristic remains robust even under weak interactions and exhibits unboundedness in infinite systems. Moreover, a clear differentiation between the ergodic and MBL phases is achieved by analyzing the quantum coherence dynamics. In particular, in the ergodic phase, the local coherence rapidly decays to its minimum, while the total coherence rapidly increases to its maximum, and the local coherence cannot be preserved.

Quantum coherence can be used to measure both the local coherence of specific subsystems and the total coherence of the whole system in a consistent manner.  We investigate the results of the coherent dynamics for all non-trivial subsystems and total system by initializing the system to the N\'eel state. In addition, we compare the results obtained from different initial states and different measures of quantum coherence to verify the validity in detecting MBL features.

In particular, exploiting the dynamics of quantum coherence as a probe to detect MBL offers advantages over entanglement under current experimental conditions. First, when a maximally coherent state is employed as an initial state, single-site local coherence can effectively distinguish between MBL and AL, unlike entanglement, which requires the joint detection of at least two sites.  Therefore, it is more feasible to implement it experimentally. In addition, direct measurement of quantum coherence via interference fringes and comparison with $l_1$-norm quantum coherence obtained via quantum state tomography yielded completely consistent results~\cite{PhysRevLett.118.020403}. Moreover, other experimental approaches to quantifying quantum coherence~\cite{PhysRevLett.113.170401,Hou2020Direct} collectively demonstrate that achieving quantum coherence is more attainable than quantum entanglement under existing experimental conditions.

\acknowledgments

This work was supported by the National Natural Science Foundation of China (Grant Nos. 12004280, 12204348, 12105074, 92265207) and the Natural Science Foundation of Guangdong Province (Grant No. 2024A1515010398).

\appendix

\section{Ergodic-MBL transition point from level statistics}

\begin{figure}[htbp]
  \includegraphics[width=0.98\columnwidth]{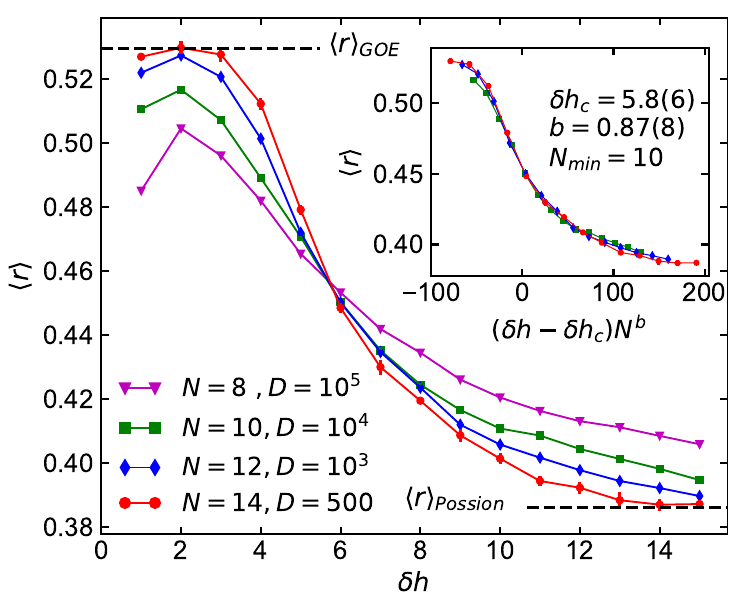}
  \caption{The average level spacing ratio $\langle r \rangle$  as a function of disorder strength $\delta h$ for different system sizes $N$ in the half-filled sector. The number of disorder realizations employed is $10^5,10^4,10^3$, and $500$ for $N = 8,10,12,14$, respectively. The inset shows the data collapse using the best estimates for the critical disorder strength $\delta h_c=5.8(6)$ and $b=0.87(8)$, excluding the data for $\delta h=1$ and $N<N_{min}$ to ensure the stability of the results.}
  \label{LevelStatistics:fig}
\end{figure}

The level statistics provide a powerful and basis-independent diagnostic to distinguish between the ergodic and localized phase~\cite{PhysRevB.82.174411,PhysRevB.93.041424,PhysRevB.97.035163,PhysRevLett.122.180601}. For our model with the Hamiltonian in Eq.~\eqref{eq:Hamiltonian}, we performed the exact diagonalization and calculate the energy level spacing $\delta_{n} = |E_{n} - E_{n-1}|$ in the half-filled sector, where $E_{n}$ is the eigenenergy for the $n$-th eigenstate. We then examine the ratio of adjacent energy level spacings, $r_{n} = \mbox{min}\{\delta_{n},\delta_{n+1}\} / \mbox{max}\{\delta_{n},\delta_{n+1}\}$  and average this quantity over disorder realizations to obtain $\langle r \rangle$, as shown in Fig.~\ref{LevelStatistics:fig}. According to random matrix theory, in the ergodic phase, the level spacings follow the Gaussian Orthogonal Ensemble (GOE) with an average ratio of $\langle r \rangle_{\rm GOE} \approx 0.529 $. In the localized phase, they follow a Poisson distribution with $\langle r \rangle_{\rm Poisson} \approx 0.386$, for $N \rightarrow \infty$. Note that our model is integrable at $\delta h = 0$, so it does not exhibit GOE level statistics in that limit. This effect is particularly evident in our smallest system size $N = 8$ and the lowest disorder strength $\delta h = 1$ as shown in Fig.~\ref{LevelStatistics:fig}.

When varying the disorder strength $\delta h$, a clear crossing is observed. To determine the critical point, we perform a scaling collapse $g[(\delta h-\delta h_c)N^{b}]$, which enables the data to collapse onto a single universal curve. The inset yields approximately the critical disorder strength of $\delta h_c=5.8(6)$ and the critical exponent of $b=0.87(8)$.

\bibliographystyle{apsrev4-1} 

\bibliography{refs_coherence}

\end{document}